\documentclass{aastex}
\usepackage{spr-astr-addons}

\begin{document}

\title{Constraining Recoiling Velocities of Black Holes Ejected by Gravitational Radiation in Galaxy Mergers\\
{\small Accepted for publication in Astrophysics and Space Science}}

\author{M. Yu. Piotrovich} \and \author{Yu. N. Gnedin} \and \author{T. M. Natsvlishvili} \and \author{S. D. Buliga}
\affil{Central Astronomical Observatory at Pulkovo, Saint-Petersburg, Russia.}
\email{mpiotrovich@mail.ru}

\begin{abstract}
Recent general relativistic simulations have shown that the coalescence of two spinning black holes (BH) can lead to recoiling speeds of the BH remnant of up to thousands of km/s as a result of the gravitational radiation emission. It is important that the accretion disc remains bound to ejected BH within the region where the gas orbital velocity is larger than the ejection speed. We considered the situation when the recoiling kick radius coincides with the radius of the broad line region (BLR). We show that in this situation the observed polarization data of accretion disk emission allow to determine the value of the recoil velocity. We present the estimates of the kick velocity for AGN with determined polarization data.
\end{abstract}


\section{Introduction} 

Mergers of spinning BHs can produce recoil velocities (''kicks'') of the final merged BHs resulting from anisotropic gravitational radiation up to several thousands km/s \citep{loeb07,shields08,komossa08}. Gravitational wave (GW) recoil implies that the supermassive black hole (SMBH) spend a significant fraction of time off nucleus, at scales beyond that of the molecular obscurity torus. For example, according to \citet{batcheldor10}, isophotal analysis of M87, using data from the Advanced Camera for Surveys, reveals a projected displacement of $6.8 \pm 0.8$ pc ($\sim 0''.1$) between the nuclear point source (presented to be location of the SMBH) and the photo-center of the galaxy itself.

A recoiling SMBH in an active galactic nuclei (AGN) retains the inner part of its accretion disc. \citet{bonning07} and \citet{komossa08} have shown that the accretion disc will remain bound to the recoiling BH, inside the radius

\begin{equation}
 R_k = \frac{G M_{BH}}{V_k^2} = 0.43 \left(\frac{M_{BH}}{10^8 M_{\odot}}\right) \left(\frac{V_k}{10^3 \frac{km}{s}}\right)^{-2} pc,
 \label{eq00}
\end{equation}

\noindent where $V_k$ is the recoiling velocity and $M_{BH}$ is the mass of a black hole.

According to \citet{komossa08} a large fraction of the broad line region (BLR) remains bound to the recoiling hole, which structure of the size of torus or larger will typically be left behind. It means that the upper radius of the accretion disk (i.e. $R_{BLR}$ in this situation) of the recoiling BH can coincide with $R_k$ and this fact allows for the ejected BH and disk to undergo a luminous phase for observations.

\citet{bonning07} have claimed that one of possible manifestation of a recoiling accretion disc is in QSO emission lines shifted in velocity from the host galaxy and they have underlined that few, if any, of these systems are likely candidates for recoiling BHs. They have examined broad line QSOs with observed $H_\beta$ and [OIII] from the SDSS which have broad emission lines substantially shifted relative to the narrow lines. In a result they placed upper limits on the incidence of recoiling SMBHs in QSOs from the SDSS.

Gravitational recoil kicks would have a variety of observational consequences. The kick velocity of remnant depends on the mass ration ($M_2 / M_1 < 1$) and spin parameters ($a_1$, $a_2$) of the binary system, but not the total mass of the system. For example, according to \citet{bekenstein79,fitchett83,gonzalez07}, the recoil velocity for a merger of non-spinning BHs is $V_k \sim \eta^2$, where $\eta = M_1 M_2 / (M_1 + M_2)^2$. The power of gravitational waves (GW) is depending also on the parameter $\eta$. It allows to estimate the recoil velocity from the direct measurements of GWs. Early estimates of the recoil velocity, framed in post-Newtonian regime for non-spinning unequal mass BHs \citep{redmount89} yielded velocities in the range $100 < V_k < 500 km/s$. Simulations with varying (arbitrary) spin orientations \citep{campanelli07a} and mass ratios \citep{baker08} show that recoil velocities can be significantly larger, reaching $V_k \geq 2000 km/s$ and are predicted to be as large as $V_k \sim 4000 km/s$ for unequal mass ($M_2 / M_1 \approx 1/3$), maximally spinning BHs \citep{campanelli07b}. Variety of observational consequences of gravitational recoil kicks confirm the importance of future GW astronomy.

Here we suggest other probable test for the incidence of recoiling BHs. This test is connected with polarization of broad line emission in the retaining BLR. We use the theory of multiple scattering of polarized radiation \citep{chandrasekhar50,sobolev63} and the disc-like model of BLR \citep{mclure01,mclure04,kollatschny06,kollatschny13b,kollatschny13}.

Our estimates of the kick velocities do not depend on the direction of the recoiling velocity. Our basic idea is to consider the situation when the BLR radius coincides with the kick radius, i.e. $R_{BLR} \approx R_k$. This position forms a new opinion on the origin of BLR, which is still unclear. For example, we present the citation from well known paper \citet{kollatschny13b}: ''However, many details of this line emitting region are unknown, and there are many models that treat their geometry and structure.'' As concerns to the real problem of deriving the recoil velocity direction, one of the possible ways for solution of this problem is to derive the real displacement of the final SMBH and its direction.

The most important fact is that the BLR and recoiling radius can be determined from polarimetric observations \citep{piotrovich15}.

The commonly accepted method for determining the SMBH is based on the virial theorem, that is applied to the BLR. The virial theorem allows to obtain the following relation \citep{vestergaard06,fine08}:

\begin{equation}
 M_{BH} = f \frac{R_{BLR} V_{BLR}^2}{G},
 \label{eq01}
\end{equation}

\noindent where $f$ is a virial coefficient (factor) that depends strongly on the geometry, velocity field and orientation of BLR, $R_{BLR}$ is the radius of BLR and $V_{BLR}$ is the velocity dispersion that is measured usually as the full width of the emission line at a half of height in the radiation spectrum, i.e. $V_{BLR} = FWHM$. The BLR radius $R_{BLR}$ is determined usually by the reverberation method, i.e. with time delay between continuum and emission line variations.

There are various approaches for determining the value of $f$. Many authors used the value $f \approx 1.0$. \citet{peterson99} suggested the value $f = 3/4$. \citet{onken04} used the mean value of the virial coefficient as $f = 1.4$. \citet{mclure01} have shown that for a disc shaped BLR without a combination of a random isotropic components inclined at angle $i$ to the observer line of sight the virial coefficient can be presented as:

\begin{equation}
 f = \frac{1}{4 \sin^2{i}}.
 \label{eq02}
\end{equation}

For a geometrically thin disc shape BLR, Eq.(\ref{eq01}) and (\ref{eq02}) transform into

\begin{equation}
 \sin{i} = \frac{1}{2} \left(\frac{R_{BLR}}{R_g}\right)^{1/2} \left(\frac{FWHM}{c}\right),
 \label{eq03}
\end{equation}

\noindent where $FWHM$ is the observed full width of the emission line \citep{vestergaard06,ho08,wang09,shen14}, and $R_g = G M_{BH} / c^2$ is the gravitational radius.

The value of the inclination angle $i$ can be determined from polarimetric observations. We have the results of the detailed numerical calculations of the degree of polarization $P_l(\mu = \cos{i})$ for the radiation scattered in the optically thick plane parallel atmosphere. These calculations were made in the framework of the classical Chandrasekhar-Sobolev theory \citep{chandrasekhar50,sobolev63}.

\section{Relation between polarization degree of emission lines and recoiling velocity of SMBH}

We can obtain the constraints on the kick velocity value is one suggests that the next relation takes the form: $R_k \approx R_{BLR}$, If this relation occurs then we obtain from (\ref{eq02}) and (\ref{eq03}):

\begin{equation}
 \sin{i} = 0.492 \alpha \left(\frac{FWHM}{V_k}\right),\,\, \alpha = \frac{R_{BLR}}{R_k}.
 \label{eq05}
\end{equation}

\noindent We shall consider the situation when the coefficient $\alpha \leq 1$. It means that our results of $V_k$ really present bounds on the kick velocity value. It is very important that the value of the inclination angle can be determined from polarimetric observations using the standard Chandrasekhar-Sobolev theory of multiple scattering of the radiation on free electrons and Rayleigh scattering on gas molecules and small dust particles. According to these classical works, the polarization degree of scattered radiation depends strongly on the inclination angle. For example, the scattered radiation has the maximum linear polarization $P_l = 11.7\%$ when the line of sight is perpendicular to the normal to the semi infinite atmosphere (Milne problem). \citet{chandrasekhar50} and \citet{sobolev63} presented the solution of so-called Milne problem that corresponds to multiple scattering of light in optically thick flattened atmosphere.

We used the theory of multiple scattering of polarized radiation and disc like model for the BLR and estimated the values of the virial factors and the mass values for SMBH in AGN \citep{piotrovich15}.

\begin{figure}[t]
	\includegraphics[width=\columnwidth]{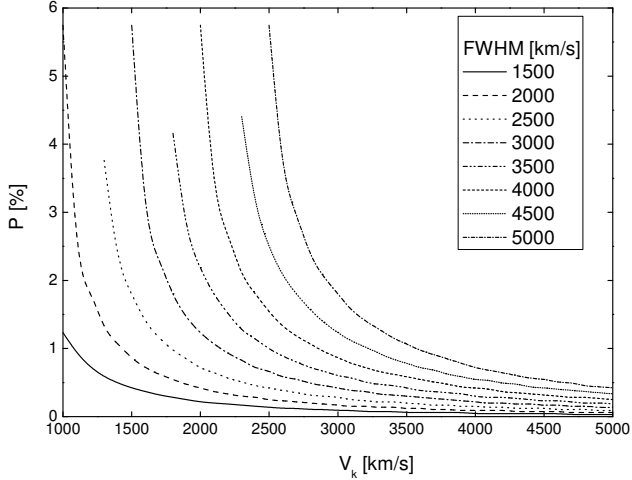}
  \caption{Dependence of the polarization degree on the kick velocity.}
  \label{fig01}
\end{figure}

Now, using Eq.(\ref{eq05}) is is possible to obtain the real relation between the degree of linear polarization and the recoil velocity $V_k$, of course, in the situation when $R_k \approx R_{BLR}$. At Fig.\ref{fig01} it is presented the dependence of the polarization degree on the kick velocity according to (\ref{eq05}) fot the different values of FWHMs.

According to \citet{lousto11} the recoil velocity of BHs with individual spin $a = 1$ has the value $V_k = 4915.2$ km/s. For $FWHM = 5 \times 10^3$ km/s the polarization degree according to (\ref{eq05}) is $P_l (H_\beta) = 0.624 \%$. The value corrsponds to the extreme kick velocity. This extreme case can only be realized by essentially particular fine tuned configuration of the individual black hole spin vectors.

\section{Estimates of the kick velocity from the spectropolarimetric data of \citet{smith02} and \citet{afanasiev11}}

For determining the kick velocity $V_k$ from Eq.(\ref{eq05}) we used the polarimetric data that are presented in the spectropolarimetric atlas of \citet{smith02}. They obtained the values of polarization degree and polarization angles for 36 Type 1 Seyfert galaxies. These data have been obtained in a result of different runs at the Anglo-Australian and William Hershel telescopes. From 36 objects presented in the atlas of \citet{smith02} for most of the observed objects there is a difference between the continuum and BLR emission values of the polarization degree and the position angle. We suggest that this difference can be connected to the contribution of the kick velocity to the broad line structure according to Eq.(\ref{eq05}). For estimates of FWHM values we used published data \citep{vestergaard06,ho08,wang09,feng14}. These data belong to the observed full width of the $H_\beta$ emission line, but the \citet{smith02} polarimetric data belong to $H_\alpha$ emission line. However existing relation between $H_\alpha$ and $H_\beta$ full width \citep{greene05} confirms the approximate equality between these widths:

\[
 FWHM(H_\beta) =
\]
\begin{equation}
 = (1.07 \pm 0.07) \times 10^3 \left(\frac{FWHM(H_\alpha)}{10^3 km/s}\right)^{1.03 \pm 0.03}
 \label{eq06}
\end{equation}

The results of our calculations of the SMBH kick velocities are presented at Table \ref{tab1}. Of course, these results can be considered as upper limits.

\begin{table}[t]
\footnotesize
\caption{Polarization of AGN radiation \citep{smith02,afanasiev11} and bounds on kick velocity values (Eq.(\ref{eq05})).}
\label{tab1}
\centering
\begin{tabular}{lccc}
\tableline
{\bf Objects} & $P_l$           & FWHM           & $V_k$ \\
              & [\%]            & [km/s]         & [km/s] \\
\tableline
Akn 564       & $0.52 \pm 0.02$ & 865            & 772 \\
I Zw 1        & $0.67 \pm 0.01$ & 1117.4         & 959.9 \\
Mrk 290       & $0.90 \pm 0.04$ & $4270 \pm 157$ & 3160 \\
Fairall 51    & $5.19 \pm 0.07$ & 3079           & 1550 \\
NGC 3783      & $0.52 \pm 0.02$ & 3555           & 3242 \\
NGC 4051      & $0.55 \pm 0.04$ & 1034           & 925 \\
Mrk 509       & $0.85 \pm 0.03$ & 3423           & 2587 \\
Mrk 335       & $0.52 \pm 0.02$ & 1840           & 1695 \\
Mrk 6         & $0.90 \pm 0.02$ & $4512 \pm 38$  & 3343.3 \\
Mrk 304       & $0.72 \pm 0.07$ & 4532           & 3637 \\
Mrk 705       & $0.46 \pm 0.07$ & 1790           & 1718 \\
Mrk 871       & $0.65 \pm 0.13$ & 3688           & 3096 \\
Mrk 876       & $0.81 \pm 0.04$ & 5017           & 3843 \\
Mrk 915       & $0.47 \pm 0.07$ & $4560 \pm 500$ & 3864 \\
NGC 4593      & $0.57 \pm 0.05$ & 3769           & 3323 \\
NGC 6814      & $1.71 \pm 0.07$ & 4200           & 2548 \\
PG 0007+106   & $1.02 \pm 0.38$ & 5084.6         & 3605 \\
PG 0026+129   & $0.99 \pm 0.28$ & 2250           & 1611 \\
PG 0049+171   & $1.42 \pm 0.31$ & 5234.3         & 3411 \\
PG 0157+001   & $0.71 \pm 0.28$ & 2432           & 1965 \\
PG 0804+761   & $1.00 \pm 0.38$ & 3276           & 2326 \\
PG 0844+349   & $0.69 \pm 0.10$ & $2694 \pm 58$  & $2194 \pm 47$ \\
PG 0953+414   & $0.39 \pm 0.12$ & $3071 \pm 27$  & $2824 \pm 28$ \\
PG 1022+519   & $0.83 \pm 0.30$ & 1566.4         & 1364 \\
PG 1116+215   & $0.46 \pm 0.10$ & 2896.9         & 2794.6 \\
PG 2112+059   & $1.14 \pm 0.22$ & 3176.4         & 2173.6 \\
PG 2130+099   & $0.62 \pm 0.15$ & $1781 \pm 5$   & $1513 \pm 4$ \\
PG 2214+139   & $1.40 \pm 0.16$ & 4532           & 2903.3 \\
PG 2233+134   & $0.67 \pm 0.23$ & 1709.2         & 1411 \\
\tableline
\end{tabular}
\end{table}

In our estimates of recoiling velocities we suggested $\alpha \leq 1$. It is interesting to compare the real value of $R_k$ from Eq.(\ref{eq00}) with values of $R_{BLR}$ published in literature. For the value of $V_k$ which is presented in Eq.(\ref{eq00}) we use the bound value of the kick velocity from our Table \ref{tab1}. The values of $R_{BLR}$, estimated by \citet{greene10,shen10,bentz13}, correspond in the error limits to values of $R_k$ from Eq.(\ref{eq00}). For example, for Fairall~9 $R_{BLR} = 10^{16.89}$cm, $R_k = 10^{16.9}$cm, for I Zw 1 $R_{BLR} = 10^{17.4}$ cm, $R_k = 10^{17.5}$cm, for Mrk~6 $R_{BLR} = 10^{16.3}$cm, $R_k = 10^{16.4}$cm, for Mrk~279 $R_{BLR} = 10^{16.6}$cm, $R_k = 10^{16.4}$cm, for Mrk~509 $R_{BLR} = 10^{17.01}$cm, $R_k = 10^{17.06}$cm, for Mrk~335 $R_{BLR} = 10^{16.64}$cm, $R_k = 10^{16.8}$cm.

For AGNs, presented in the Table \ref{tab1}, the recoil velocities are essentially lower than the upper limit for recoil $\sim 5000 km/s$. For some other AGN, including Mrk~279, Akn~120, Fairall~9, NGC~5548 and PG~2209+184 polarimetric estimates with Eqs. (\ref{eq01}) and (\ref{eq02}) give the values above this upper limits. It appears that in the case of the geometrically thick disk-like structure of BLR the polarimetric data from \citet{smith02} and \citet{afanasiev11} provide the values extremely lower the largest recoil value.

According to \citet{collin06} and \citet{decarli11} the expression for the virial factor can be presented in the following form:

\begin{equation}
 f = 0.25 \left[\left(\frac{H}{R}\right)^2 + \sin^2 i \right],\,\, V_{BLR} = FWHM,
 \label{eq07}
\end{equation}

\noindent where $H / R$ is the aspect ratio, i.e. the ratio of the geometrical thickness of the BLR to this radius ($R = R_{BLR}$). As a result we obtain instead of (\ref{eq02}) and (\ref{eq05}) the following expressions

\begin{equation}
 \sqrt{\left(\frac{H}{R}\right)^2 +\sin^2 i} = \left(\frac{FWHM}{2 c}\right) \left(\frac{R_{BLR}}{R_g}\right)^{1/2},
 \label{eq08}
\end{equation}

\begin{equation}
 \sqrt{\left(\frac{H}{R}\right)^2 + \sin^2 i} = 0.492 \alpha \left(\frac{FWHM}{V_k}\right).
 \label{eq09}
\end{equation}

The real value of the parameter $H / R$ can be obtained with (\ref{eq08}) using the values of $\sin i$ from the polarimetric observations at the base of Chandrasekhar-Sobolev theory of the generation of polarization for multiple scattering in optically thick plane-parallel atmosphere.

The values of the recoil velocities for a number of AGN, calculating with (\ref{eq07}) and (\ref{eq08}) are presented at Table \ref{tab2}.

\begin{table}[t]
\footnotesize
\caption{Polarization of AGN radiation and bounds on kick value (Eq.(\ref{eq05})) for geometrically thick accretion flow \citep{smith02,afanasiev11}.}
\label{tab2}
\centering
\begin{tabular}{lccc}
\tableline
{\bf Objects} & $P_l$           & FWHM           & $V_k$ \\
              & [\%]            & [km/s]         & [km/s] \\
\tableline
Mrk 279       & $0.48 \pm 0.04$ & $5208\pm 95$   & 2928 \\
Akn 120       & $0.40 \pm 0.02$ & $5536\pm 297$  & 2768 \\
Fairall 9     & $0.40 \pm 0.11$ & $5618\pm 107$  & 3786 \\
NGC 5548      & $0.69 \pm 0.01$ & 5822           & 3693 \\
PG 2209+184   & $0.83 \pm 0.29$ & 6487.3         & 3143 \\
\tableline
\end{tabular}
\end{table}

It should be noted that for the objects presented in the Table \ref{tab1}, the polarization degree data and corresponding values of $\sin{i}$ require the real geometrically thin BLR, i.e. $H / R \ll |\sin{i}|$.

\section{Discussion and conclusions}

According to the basic result of our calculations presented at the Table \ref{tab1} and Table \ref{tab2} the kick velocity is smaller than the FWHM value. This result confirms the conclusion of \citet{loeb07} that the accretion disc remains bound to the ejected BH within the region where the gas orbital velocity is larger than ejection speed.

The main problem is the fact that the remnant BH recoiling in any direction can exceed the escape velocity of galaxies. In this situation the recoil velocity must essentially exceed $\sim 2000$ km/s.

According to \citet{loeb07} only the small fraction of quasars could be associated with an escaping BH. If the accreting disk remains bound to ejected BH, one can expect the radiation of this disk would be polarized. Our estimates of the recoiling velocities correspond to the situation when $R_{BLR} \approx R_k$. But in many cases the situation is realized when $R_{BLR} < R_k$ and in these cases the values of $V_k$ will be considerably lower than obtained values. Therefore our results can be considered as bounds for the real velocity values.

Also it is important that according to \citet{peterson07} and \citet{komossa12} the region of the kick velocity ($R_k$ from Eq.(\ref{eq00})) is on the order of the size of the BLR of AGN.

Numerical relativistic simulations have produced recoil velocities $V_k \geq 10^3$ km/s, even reaching $\sim 5 \times 10^3$ km/s for certain configurations of BH spins \citep{lena14,campanelli07b,lousto11,lousto12}. Such velocities would cause large displacements of the coalesced SMBH from the center of galaxy or, in the extreme cases, eject it entirely from the host galaxy \citep{merritt04,volonteri10}. But recoils exceeding the escape velocity of the host galaxy are expected to be relatively rare. It is more important that coalesced SMBH will undergo damped oscillations that will prevent the real escape of the object from the host galaxy.

\citet{merritt09} and \citet{gualandris08} have shown that one of the key consequences of GW recoil is long lasting oscillations of the SMBH around the host galaxy core, implying that the SMBHs may spend as long as $10^6 \div 10^9$ years of the nucleus with an amplitude of parsecs or kiloparsecs. For example, N-body simulations have shown that the SMBH can oscillate within the bulge for $\sim 1$ Gyr before coming to rest \citep{gualandris08}.

Also it is important that recoiling BH trajectories strongly depend on the gas content of the host galaxy. Maximal BH displacements from the center may vary by up to an order of magnitude between gas rich and gas poor mergers.

The problem of oscillating of BH kicks in host galaxies is the extremely complex and deserves the special investigation. The base of our consideration is the suggestion that the radius of BLR coincides with the initial kick radius determined by Eq.(\ref{eq03}) \citep{komossa08}. If even the intrinsic BH velocity oscillate, the initial BLR remains bound to the recoiling hole and therefore keeps the information on the initial recoiling velocity.

It should be mentioned that last investigations showed that the probability that a remnant BH recoils in any direction at a velocity exceeding $\sim 2000 km/s$ (escape velocity of large elliptical galaxies) in only 0.03\% \citep{lousto12}.

For all AGNs presented at our Table \ref{tab1} and Table \ref{tab2} the kick velocity appears smaller than the FWHM value though reach the values of thousands km/s. Of course, these values should be considered as the limit value. It is very important that our estimates confirm that a recoiling SMBH retained the inner part of its accretion disc and in many cases $R_k \approx R_{BLR}$.

Recently \citet{robinson10} have presented the results of spectropolarimetric observations of the quasar E1821+643 ($z = 0.297$). For this object the broad Balmer lines in total flux are redshifted by $\sim 10^3$ km/s relative to the narrow lines.

At first sight the quite large values of the recoiling velocity, presented at Table \ref{tab1}, can exceed the escape velocity of the host galaxy. But, according to \citet{lousto12}, these events are expected to be relatively rare. More frequently the SMBH will undergo damped oscillations in the galaxy potential. N-body simulations \citep{gualandris08} have shown that moderately large kicks, that can be sufficient to eject the SMBH from the core, result in long lived oscillations which damp on a timescale $\sim 1$ Gyr.

Systematic searches of large SDSS AGN samples have revealed $\sim 100$ objects that exhibit velocity shifts $\geq 10^3$ km/s between the broad and narrow lines \citep{bonning07,tsalmantza11,eracleous12}. We intend to arrange the special polarimetric programm of observations of these objects at the Russian BTA-6m telescope.

\acknowledgments

This research was supported by the Basic Research Program P-7 of Praesidium of Russian Academy od Sciences, the program of the Department of Physical Sciences of Russian Academy of Sciences No.2 and the presidential program ''Leading Scientific School-7241.2016.2''.

\sloppy
\bibliographystyle{spr-mp-nameyear-cnd}
\bibliography{mybibfile}

\begin{thebibliography}{48}
\ifx \bisbn   \undefined \def \bisbn  #1{ISBN #1}\fi
\ifx \binits  \undefined \def \binits#1{#1} \fi
\ifx \bauthor  \undefined \def \bauthor#1{#1} \fi
\ifx \batitle  \undefined \def \batitle#1{#1} \fi
\ifx \bjtitle  \undefined \def \bjtitle#1{#1}\fi
\ifx \bvolume  \undefined \def \bvolume#1{\textbf{#1}}\fi
\ifx \byear  \undefined \def \byear#1{#1} \fi
\ifx \bissue  \undefined \def \bissue#1{#1} \fi
\ifx \bfpage  \undefined \def \bfpage#1{#1} \fi
\ifx \blpage  \undefined \def \blpage #1{#1} \fi
\ifx \burl  \undefined \def \burl#1{\textsf{#1}} \fi
\ifx \doiurl  \undefined \def \doiurl#1{\textsf{#1}} \fi
\ifx \betal  \undefined \def \betal{\textit{et al.}} \fi
\ifx \binstitute  \undefined \def \binstitute#1{#1} \fi
\ifx \binstitutionaled  \undefined \def \binstitutionaled#1{#1} \fi
\ifx \bctitle  \undefined \def \bctitle#1{#1} \fi
\ifx \beditor  \undefined \def \beditor#1{#1} \fi
\ifx \bpublisher  \undefined \def \bpublisher#1{#1} \fi
\ifx \bbtitle  \undefined \def \bbtitle#1{#1} \fi
\ifx \bedition  \undefined \def \bedition#1{#1} \fi
\ifx \bseriesno  \undefined \def \bseriesno#1{#1} \fi
\ifx \blocation  \undefined \def \blocation#1{#1} \fi
\ifx \bsertitle  \undefined \def \bsertitle#1{#1} \fi
\ifx \bsnm \undefined \def \bsnm#1{#1} \fi
\ifx \bsuffix \undefined \def \bsuffix#1{#1} \fi
\ifx \bparticle \undefined \def \bparticle#1{#1} \fi
\ifx \barticle \undefined \def \barticle#1{#1} \fi
\ifx \bconfdate \undefined \def \bconfdate #1{#1} \fi
\ifx \botherref \undefined \def \botherref #1{#1} \fi
\ifx \url \undefined \def \url#1{\textsf{#1}} \fi
\ifx \bchapter \undefined \def \bchapter#1{#1} \fi
\ifx \bbook \undefined \def \bbook#1{#1} \fi
\ifx \bcomment \undefined \def \bcomment#1{#1} \fi
\ifx \oauthor \undefined \def \oauthor#1{#1} \fi
\ifx \citeauthoryear \undefined \def \citeauthoryear#1{#1} \fi
\ifx \endbibitem  \undefined \def \endbibitem {}\fi
\ifx \bconflocation  \undefined \def \bconflocation#1{#1} \fi
\ifx \arxivurl  \undefined \def \arxivurl#1{\textsf{#1}} \fi

\bibitem[\protect\citeauthoryear{{Afanasiev} et~al.}{2011}]{afanasiev11}
\begin{barticle}
\bauthor{\bsnm{{Afanasiev}}, \binits{V.L.}},
\bauthor{\bsnm{{Borisov}}, \binits{N.V.}},
\bauthor{\bsnm{{Gnedin}}, \binits{Y.N.}},
\bauthor{\bsnm{{Natsvlishvili}}, \binits{T.M.}},
\bauthor{\bsnm{{Piotrovich}}, \binits{M.Y.}},
\bauthor{\bsnm{{Buliga}}, \binits{S.D.}}:
\bjtitle{Astronomy Letters}
\bvolume{37},
\bfpage{302}
(\byear{2011}).
\arxivurl{1104.3690}.
doi:\doiurl{10.1134/S106377371105001X}
\end{barticle}
\endbibitem

\bibitem[\protect\citeauthoryear{{Baker} et~al.}{2008}]{baker08}
\begin{barticle}
\bauthor{\bsnm{{Baker}}, \binits{J.G.}},
\bauthor{\bsnm{{Boggs}}, \binits{W.D.}},
\bauthor{\bsnm{{Centrella}}, \binits{J.}},
\bauthor{\bsnm{{Kelly}}, \binits{B.J.}},
\bauthor{\bsnm{{McWilliams}}, \binits{S.T.}},
\bauthor{\bsnm{{Miller}}, \binits{M.C.}},
\bauthor{\bsnm{{van Meter}}, \binits{J.R.}}:
\bjtitle{\apjl}
\bvolume{682},
\bfpage{29}
(\byear{2008}).
\arxivurl{0802.0416}.
doi:\doiurl{10.1086/590927}
\end{barticle}
\endbibitem

\bibitem[\protect\citeauthoryear{{Batcheldor} et~al.}{2010}]{batcheldor10}
\begin{barticle}
\bauthor{\bsnm{{Batcheldor}}, \binits{D.}},
\bauthor{\bsnm{{Robinson}}, \binits{A.}},
\bauthor{\bsnm{{Axon}}, \binits{D.J.}},
\bauthor{\bsnm{{Perlman}}, \binits{E.S.}},
\bauthor{\bsnm{{Merritt}}, \binits{D.}}:
\bjtitle{\apjl}
\bvolume{717},
\bfpage{6}
(\byear{2010}).
\arxivurl{1005.2173}.
doi:\doiurl{10.1088/2041-8205/717/1/L6}
\end{barticle}
\endbibitem

\bibitem[\protect\citeauthoryear{{Bekenstein}}{1979}]{bekenstein79}
\begin{barticle}
\bauthor{\bsnm{{Bekenstein}}, \binits{J.D.}}:
\bjtitle{Comments on Astrophysics}
\bvolume{8},
\bfpage{89}
(\byear{1979})
\end{barticle}
\endbibitem

\bibitem[\protect\citeauthoryear{{Bentz} et~al.}{2013}]{bentz13}
\begin{barticle}
\bauthor{\bsnm{{Bentz}}, \binits{M.C.}},
\bauthor{\bsnm{{Denney}}, \binits{K.D.}},
\bauthor{\bsnm{{Grier}}, \binits{C.J.}},
\bauthor{\bsnm{{Barth}}, \binits{A.J.}},
\bauthor{\bsnm{{Peterson}}, \binits{B.M.}},
\bauthor{\bsnm{{Vestergaard}}, \binits{M.}},
\bauthor{\bsnm{{Bennert}}, \binits{V.N.}},
\bauthor{\bsnm{{Canalizo}}, \binits{G.}},
\bauthor{\bsnm{{De Rosa}}, \binits{G.}},
\bauthor{\bsnm{{Filippenko}}, \binits{A.V.}},
\bauthor{\bsnm{{Gates}}, \binits{E.L.}},
\bauthor{\bsnm{{Greene}}, \binits{J.E.}},
\bauthor{\bsnm{{Li}}, \binits{W.}},
\bauthor{\bsnm{{Malkan}}, \binits{M.A.}},
\bauthor{\bsnm{{Pogge}}, \binits{R.W.}},
\bauthor{\bsnm{{Stern}}, \binits{D.}},
\bauthor{\bsnm{{Treu}}, \binits{T.}},
\bauthor{\bsnm{{Woo}}, \binits{J.-H.}}:
\bjtitle{\apj}
\bvolume{767},
\bfpage{149}
(\byear{2013}).
\arxivurl{1303.1742}.
doi:\doiurl{10.1088/0004-637X/767/2/149}
\end{barticle}
\endbibitem

\bibitem[\protect\citeauthoryear{{Bonning} et~al.}{2007}]{bonning07}
\begin{barticle}
\bauthor{\bsnm{{Bonning}}, \binits{E.W.}},
\bauthor{\bsnm{{Shields}}, \binits{G.A.}},
\bauthor{\bsnm{{Salviander}}, \binits{S.}}:
\bjtitle{\apjl}
\bvolume{666},
\bfpage{13}
(\byear{2007}).
\arxivurl{0705.4263}.
doi:\doiurl{10.1086/521674}
\end{barticle}
\endbibitem

\bibitem[\protect\citeauthoryear{{Campanelli} et~al.}{2007a}]{campanelli07a}
\begin{barticle}
\bauthor{\bsnm{{Campanelli}}, \binits{M.}},
\bauthor{\bsnm{{Lousto}}, \binits{C.}},
\bauthor{\bsnm{{Zlochower}}, \binits{Y.}},
\bauthor{\bsnm{{Merritt}}, \binits{D.}}:
\bjtitle{\apjl}
\bvolume{659},
\bfpage{5}
(\byear{2007}a).
\arxivurl{gr-qc/0701164}.
doi:\doiurl{10.1086/516712}
\end{barticle}
\endbibitem

\bibitem[\protect\citeauthoryear{{Campanelli} et~al.}{2007b}]{campanelli07b}
\begin{barticle}
\bauthor{\bsnm{{Campanelli}}, \binits{M.}},
\bauthor{\bsnm{{Lousto}}, \binits{C.O.}},
\bauthor{\bsnm{{Zlochower}}, \binits{Y.}},
\bauthor{\bsnm{{Merritt}}, \binits{D.}}:
\bjtitle{Physical Review Letters}
\bvolume{98}(\bissue{23}),
\bfpage{231102}
(\byear{2007}b).
\arxivurl{gr-qc/0702133}.
doi:\doiurl{10.1103/PhysRevLett.98.231102}
\end{barticle}
\endbibitem

\bibitem[\protect\citeauthoryear{{Chandrasekhar}}{1950}]{chandrasekhar50}
\begin{bbook}
\bauthor{\bsnm{{Chandrasekhar}}, \binits{S.}}:
\bbtitle{{Radiative Transfer.}},
(\byear{1950})
\end{bbook}
\endbibitem

\bibitem[\protect\citeauthoryear{{Collin} et~al.}{2006}]{collin06}
\begin{barticle}
\bauthor{\bsnm{{Collin}}, \binits{S.}},
\bauthor{\bsnm{{Kawaguchi}}, \binits{T.}},
\bauthor{\bsnm{{Peterson}}, \binits{B.M.}},
\bauthor{\bsnm{{Vestergaard}}, \binits{M.}}:
\bjtitle{\aap}
\bvolume{456},
\bfpage{75}
(\byear{2006}).
\arxivurl{astro-ph/0603460}.
doi:\doiurl{10.1051/0004-6361:20064878}
\end{barticle}
\endbibitem

\bibitem[\protect\citeauthoryear{{Decarli} et~al.}{2011}]{decarli11}
\begin{barticle}
\bauthor{\bsnm{{Decarli}}, \binits{R.}},
\bauthor{\bsnm{{Dotti}}, \binits{M.}},
\bauthor{\bsnm{{Treves}}, \binits{A.}}:
\bjtitle{\mnras}
\bvolume{413},
\bfpage{39}
(\byear{2011}).
\arxivurl{1011.5879}.
doi:\doiurl{10.1111/j.1365-2966.2010.18102.x}
\end{barticle}
\endbibitem

\bibitem[\protect\citeauthoryear{{Eracleous} et~al.}{2012}]{eracleous12}
\begin{barticle}
\bauthor{\bsnm{{Eracleous}}, \binits{M.}},
\bauthor{\bsnm{{Boroson}}, \binits{T.A.}},
\bauthor{\bsnm{{Halpern}}, \binits{J.P.}},
\bauthor{\bsnm{{Liu}}, \binits{J.}}:
\bjtitle{\apjs}
\bvolume{201},
\bfpage{23}
(\byear{2012}).
\arxivurl{1106.2952}.
doi:\doiurl{10.1088/0067-0049/201/2/23}
\end{barticle}
\endbibitem

\bibitem[\protect\citeauthoryear{{Feng} et~al.}{2014}]{feng14}
\begin{barticle}
\bauthor{\bsnm{{Feng}}, \binits{H.}},
\bauthor{\bsnm{{Shen}}, \binits{Y.}},
\bauthor{\bsnm{{Li}}, \binits{H.}}:
\bjtitle{\apj}
\bvolume{794},
\bfpage{77}
(\byear{2014}).
\arxivurl{1408.6952}.
doi:\doiurl{10.1088/0004-637X/794/1/77}
\end{barticle}
\endbibitem

\bibitem[\protect\citeauthoryear{{Fine} et~al.}{2008}]{fine08}
\begin{barticle}
\bauthor{\bsnm{{Fine}}, \binits{S.}},
\bauthor{\bsnm{{Croom}}, \binits{S.M.}},
\bauthor{\bsnm{{Hopkins}}, \binits{P.F.}},
\bauthor{\bsnm{{Hernquist}}, \binits{L.}},
\bauthor{\bsnm{{Bland-Hawthorn}}, \binits{J.}},
\bauthor{\bsnm{{Colless}}, \binits{M.}},
\bauthor{\bsnm{{Hall}}, \binits{P.B.}},
\bauthor{\bsnm{{Miller}}, \binits{L.}},
\bauthor{\bsnm{{Myers}}, \binits{A.D.}},
\bauthor{\bsnm{{Nichol}}, \binits{R.}},
\bauthor{\bsnm{{Pimbblet}}, \binits{K.A.}},
\bauthor{\bsnm{{Ross}}, \binits{N.P.}},
\bauthor{\bsnm{{Schneider}}, \binits{D.P.}},
\bauthor{\bsnm{{Shanks}}, \binits{T.}},
\bauthor{\bsnm{{Sharp}}, \binits{R.G.}}:
\bjtitle{\mnras}
\bvolume{390},
\bfpage{1413}
(\byear{2008}).
\arxivurl{0807.1155}.
doi:\doiurl{10.1111/j.1365-2966.2008.13691.x}
\end{barticle}
\endbibitem

\bibitem[\protect\citeauthoryear{{Fitchett}}{1983}]{fitchett83}
\begin{barticle}
\bauthor{\bsnm{{Fitchett}}, \binits{M.J.}}:
\bjtitle{\mnras}
\bvolume{203},
\bfpage{1049}
(\byear{1983}).
doi:\doiurl{10.1093/mnras/203.4.1049}
\end{barticle}
\endbibitem

\bibitem[\protect\citeauthoryear{{Gonz{\'a}lez} et~al.}{2007}]{gonzalez07}
\begin{barticle}
\bauthor{\bsnm{{Gonz{\'a}lez}}, \binits{J.A.}},
\bauthor{\bsnm{{Hannam}}, \binits{M.}},
\bauthor{\bsnm{{Sperhake}}, \binits{U.}},
\bauthor{\bsnm{{Br{\"u}gmann}}, \binits{B.}},
\bauthor{\bsnm{{Husa}}, \binits{S.}}:
\bjtitle{Physical Review Letters}
\bvolume{98}(\bissue{23}),
\bfpage{231101}
(\byear{2007}).
\arxivurl{gr-qc/0702052}.
doi:\doiurl{10.1103/PhysRevLett.98.231101}
\end{barticle}
\endbibitem

\bibitem[\protect\citeauthoryear{{Greene} and {Ho}}{2005}]{greene05}
\begin{barticle}
\bauthor{\bsnm{{Greene}}, \binits{J.E.}},
\bauthor{\bsnm{{Ho}}, \binits{L.C.}}:
\bjtitle{\apj}
\bvolume{630},
\bfpage{122}
(\byear{2005}).
\arxivurl{astro-ph/0508335}.
doi:\doiurl{10.1086/431897}
\end{barticle}
\endbibitem

\bibitem[\protect\citeauthoryear{{Greene} et~al.}{2010}]{greene10}
\begin{barticle}
\bauthor{\bsnm{{Greene}}, \binits{J.E.}},
\bauthor{\bsnm{{Hood}}, \binits{C.E.}},
\bauthor{\bsnm{{Barth}}, \binits{A.J.}},
\bauthor{\bsnm{{Bennert}}, \binits{V.N.}},
\bauthor{\bsnm{{Bentz}}, \binits{M.C.}},
\bauthor{\bsnm{{Filippenko}}, \binits{A.V.}},
\bauthor{\bsnm{{Gates}}, \binits{E.}},
\bauthor{\bsnm{{Malkan}}, \binits{M.A.}},
\bauthor{\bsnm{{Treu}}, \binits{T.}},
\bauthor{\bsnm{{Walsh}}, \binits{J.L.}},
\bauthor{\bsnm{{Woo}}, \binits{J.-H.}}:
\bjtitle{\apj}
\bvolume{723},
\bfpage{409}
(\byear{2010}).
\arxivurl{1009.0532}.
doi:\doiurl{10.1088/0004-637X/723/1/409}
\end{barticle}
\endbibitem

\bibitem[\protect\citeauthoryear{{Gualandris} and
  {Merritt}}{2008}]{gualandris08}
\begin{barticle}
\bauthor{\bsnm{{Gualandris}}, \binits{A.}},
\bauthor{\bsnm{{Merritt}}, \binits{D.}}:
\bjtitle{\apj}
\bvolume{678},
\bfpage{780}
(\byear{2008}).
\arxivurl{0708.0771}.
doi:\doiurl{10.1086/586877}
\end{barticle}
\endbibitem

\bibitem[\protect\citeauthoryear{{Ho} et~al.}{2008}]{ho08}
\begin{barticle}
\bauthor{\bsnm{{Ho}}, \binits{L.C.}},
\bauthor{\bsnm{{Darling}}, \binits{J.}},
\bauthor{\bsnm{{Greene}}, \binits{J.E.}}:
\bjtitle{\apjs}
\bvolume{177},
\bfpage{103}
(\byear{2008}).
\arxivurl{0803.2023}.
doi:\doiurl{10.1086/588217}
\end{barticle}
\endbibitem

\bibitem[\protect\citeauthoryear{{Kollatschny} and
  {Zetzl}}{2013a}]{kollatschny13b}
\begin{barticle}
\bauthor{\bsnm{{Kollatschny}}, \binits{W.}},
\bauthor{\bsnm{{Zetzl}}, \binits{M.}}:
\bjtitle{\aap}
\bvolume{549},
\bfpage{100}
(\byear{2013}a).
\arxivurl{1211.3065}.
doi:\doiurl{10.1051/0004-6361/201219411}
\end{barticle}
\endbibitem

\bibitem[\protect\citeauthoryear{{Kollatschny} and
  {Zetzl}}{2013b}]{kollatschny13}
\begin{barticle}
\bauthor{\bsnm{{Kollatschny}}, \binits{W.}},
\bauthor{\bsnm{{Zetzl}}, \binits{M.}}:
\bjtitle{\aap}
\bvolume{558},
\bfpage{26}
(\byear{2013}b).
\arxivurl{1308.1902}.
doi:\doiurl{10.1051/0004-6361/201321685}
\end{barticle}
\endbibitem

\bibitem[\protect\citeauthoryear{{Kollatschny} et~al.}{2006}]{kollatschny06}
\begin{barticle}
\bauthor{\bsnm{{Kollatschny}}, \binits{W.}},
\bauthor{\bsnm{{Zetzl}}, \binits{M.}},
\bauthor{\bsnm{{Dietrich}}, \binits{M.}}:
\bjtitle{\aap}
\bvolume{454},
\bfpage{459}
(\byear{2006}).
doi:\doiurl{10.1051/0004-6361:20054357}
\end{barticle}
\endbibitem

\bibitem[\protect\citeauthoryear{{Komossa}}{2012}]{komossa12}
\begin{barticle}
\bauthor{\bsnm{{Komossa}}, \binits{S.}}:
\bjtitle{Advances in Astronomy}
\bvolume{2012},
\bfpage{364973}
(\byear{2012}).
\arxivurl{1202.1977}.
doi:\doiurl{10.1155/2012/364973}
\end{barticle}
\endbibitem

\bibitem[\protect\citeauthoryear{{Komossa} and {Merritt}}{2008}]{komossa08}
\begin{barticle}
\bauthor{\bsnm{{Komossa}}, \binits{S.}},
\bauthor{\bsnm{{Merritt}}, \binits{D.}}:
\bjtitle{\apjl}
\bvolume{689},
\bfpage{89}
(\byear{2008}).
\arxivurl{0811.1037}.
doi:\doiurl{10.1086/595883}
\end{barticle}
\endbibitem

\bibitem[\protect\citeauthoryear{{Lena} et~al.}{2014}]{lena14}
\begin{barticle}
\bauthor{\bsnm{{Lena}}, \binits{D.}},
\bauthor{\bsnm{{Robinson}}, \binits{A.}},
\bauthor{\bsnm{{Marconi}}, \binits{A.}},
\bauthor{\bsnm{{Axon}}, \binits{D.J.}},
\bauthor{\bsnm{{Capetti}}, \binits{A.}},
\bauthor{\bsnm{{Merritt}}, \binits{D.}},
\bauthor{\bsnm{{Batcheldor}}, \binits{D.}}:
\bjtitle{\apj}
\bvolume{795},
\bfpage{146}
(\byear{2014}).
\arxivurl{1409.3976}.
doi:\doiurl{10.1088/0004-637X/795/2/146}
\end{barticle}
\endbibitem

\bibitem[\protect\citeauthoryear{{Loeb}}{2007}]{loeb07}
\begin{barticle}
\bauthor{\bsnm{{Loeb}}, \binits{A.}}:
\bjtitle{Physical Review Letters}
\bvolume{99}(\bissue{4}),
\bfpage{041103}
(\byear{2007}).
\arxivurl{astro-ph/0703722}.
doi:\doiurl{10.1103/PhysRevLett.99.041103}
\end{barticle}
\endbibitem

\bibitem[\protect\citeauthoryear{{Lousto} and {Zlochower}}{2011}]{lousto11}
\begin{barticle}
\bauthor{\bsnm{{Lousto}}, \binits{C.O.}},
\bauthor{\bsnm{{Zlochower}}, \binits{Y.}}:
\bjtitle{Physical Review Letters}
\bvolume{107}(\bissue{23}),
\bfpage{231102}
(\byear{2011}).
\arxivurl{1108.2009}.
doi:\doiurl{10.1103/PhysRevLett.107.231102}
\end{barticle}
\endbibitem

\bibitem[\protect\citeauthoryear{{Lousto} et~al.}{2012}]{lousto12}
\begin{barticle}
\bauthor{\bsnm{{Lousto}}, \binits{C.O.}},
\bauthor{\bsnm{{Zlochower}}, \binits{Y.}},
\bauthor{\bsnm{{Dotti}}, \binits{M.}},
\bauthor{\bsnm{{Volonteri}}, \binits{M.}}:
\bjtitle{\prd}
\bvolume{85}(\bissue{8}),
\bfpage{084015}
(\byear{2012}).
\arxivurl{1201.1923}.
doi:\doiurl{10.1103/PhysRevD.85.084015}
\end{barticle}
\endbibitem

\bibitem[\protect\citeauthoryear{{McLure} and {Dunlop}}{2001}]{mclure01}
\begin{barticle}
\bauthor{\bsnm{{McLure}}, \binits{R.J.}},
\bauthor{\bsnm{{Dunlop}}, \binits{J.S.}}:
\bjtitle{\mnras}
\bvolume{327},
\bfpage{199}
(\byear{2001}).
\arxivurl{astro-ph/0009406}.
doi:\doiurl{10.1046/j.1365-8711.2001.04709.x}
\end{barticle}
\endbibitem

\bibitem[\protect\citeauthoryear{{McLure} and {Dunlop}}{2004}]{mclure04}
\begin{barticle}
\bauthor{\bsnm{{McLure}}, \binits{R.J.}},
\bauthor{\bsnm{{Dunlop}}, \binits{J.S.}}:
\bjtitle{\mnras}
\bvolume{352},
\bfpage{1390}
(\byear{2004}).
\arxivurl{astro-ph/0310267}.
doi:\doiurl{10.1111/j.1365-2966.2004.08034.x}
\end{barticle}
\endbibitem

\bibitem[\protect\citeauthoryear{{Merritt} et~al.}{2009}]{merritt09}
\begin{barticle}
\bauthor{\bsnm{{Merritt}}, \binits{D.}},
\bauthor{\bsnm{{Schnittman}}, \binits{J.D.}},
\bauthor{\bsnm{{Komossa}}, \binits{S.}}:
\bjtitle{\apj}
\bvolume{699},
\bfpage{1690}
(\byear{2009}).
\arxivurl{0809.5046}.
doi:\doiurl{10.1088/0004-637X/699/2/1690}
\end{barticle}
\endbibitem

\bibitem[\protect\citeauthoryear{{Merritt} et~al.}{2004}]{merritt04}
\begin{barticle}
\bauthor{\bsnm{{Merritt}}, \binits{D.}},
\bauthor{\bsnm{{Milosavljevi{\'c}}}, \binits{M.}},
\bauthor{\bsnm{{Favata}}, \binits{M.}},
\bauthor{\bsnm{{Hughes}}, \binits{S.A.}},
\bauthor{\bsnm{{Holz}}, \binits{D.E.}}:
\bjtitle{\apjl}
\bvolume{607},
\bfpage{9}
(\byear{2004}).
\arxivurl{astro-ph/0402057}.
doi:\doiurl{10.1086/421551}
\end{barticle}
\endbibitem

\bibitem[\protect\citeauthoryear{{Onken} et~al.}{2004}]{onken04}
\begin{barticle}
\bauthor{\bsnm{{Onken}}, \binits{C.A.}},
\bauthor{\bsnm{{Ferrarese}}, \binits{L.}},
\bauthor{\bsnm{{Merritt}}, \binits{D.}},
\bauthor{\bsnm{{Peterson}}, \binits{B.M.}},
\bauthor{\bsnm{{Pogge}}, \binits{R.W.}},
\bauthor{\bsnm{{Vestergaard}}, \binits{M.}},
\bauthor{\bsnm{{Wandel}}, \binits{A.}}:
\bjtitle{\apj}
\bvolume{615},
\bfpage{645}
(\byear{2004}).
\arxivurl{astro-ph/0407297}.
doi:\doiurl{10.1086/424655}
\end{barticle}
\endbibitem

\bibitem[\protect\citeauthoryear{{Peterson}}{2007}]{peterson07}
\begin{bchapter}
\bauthor{\bsnm{{Peterson}}, \binits{B.M.}}:
In: \beditor{\bsnm{{Ho}}, \binits{L.C.}},
\beditor{\bsnm{{Wang}}, \binits{J.-W.}} (eds.)
\bbtitle{The Central Engine of Active Galactic Nuclei}.
\bsertitle{Astronomical Society of the Pacific Conference Series},
vol. \bseriesno{373},
p. \bfpage{3}
(\byear{2007}).
\arxivurl{astro-ph/0703197}
\end{bchapter}
\endbibitem

\bibitem[\protect\citeauthoryear{{Peterson} and {Wandel}}{1999}]{peterson99}
\begin{barticle}
\bauthor{\bsnm{{Peterson}}, \binits{B.M.}},
\bauthor{\bsnm{{Wandel}}, \binits{A.}}:
\bjtitle{\apjl}
\bvolume{521},
\bfpage{95}
(\byear{1999}).
\arxivurl{astro-ph/9905382}.
doi:\doiurl{10.1086/312190}
\end{barticle}
\endbibitem

\bibitem[\protect\citeauthoryear{{Piotrovich} et~al.}{2015}]{piotrovich15}
\begin{barticle}
\bauthor{\bsnm{{Piotrovich}}, \binits{M.Y.}},
\bauthor{\bsnm{{Gnedin}}, \binits{Y.N.}},
\bauthor{\bsnm{{Silant'ev}}, \binits{N.A.}},
\bauthor{\bsnm{{Natsvlishvili}}, \binits{T.M.}},
\bauthor{\bsnm{{Buliga}}, \binits{S.D.}}:
\bjtitle{\mnras}
\bvolume{454},
\bfpage{1157}
(\byear{2015}).
\arxivurl{1509.01028}.
doi:\doiurl{10.1093/mnras/stv2047}
\end{barticle}
\endbibitem

\bibitem[\protect\citeauthoryear{{Redmount} and {Rees}}{1989}]{redmount89}
\begin{barticle}
\bauthor{\bsnm{{Redmount}}, \binits{I.H.}},
\bauthor{\bsnm{{Rees}}, \binits{M.J.}}:
\bjtitle{Comments on Astrophysics}
\bvolume{14},
\bfpage{165}
(\byear{1989})
\end{barticle}
\endbibitem

\bibitem[\protect\citeauthoryear{{Robinson} et~al.}{2010}]{robinson10}
\begin{barticle}
\bauthor{\bsnm{{Robinson}}, \binits{A.}},
\bauthor{\bsnm{{Young}}, \binits{S.}},
\bauthor{\bsnm{{Axon}}, \binits{D.J.}},
\bauthor{\bsnm{{Kharb}}, \binits{P.}},
\bauthor{\bsnm{{Smith}}, \binits{J.E.}}:
\bjtitle{\apjl}
\bvolume{717},
\bfpage{122}
(\byear{2010}).
\arxivurl{1006.0993}.
doi:\doiurl{10.1088/2041-8205/717/2/L122}
\end{barticle}
\endbibitem

\bibitem[\protect\citeauthoryear{{Shen} and {Ho}}{2014}]{shen14}
\begin{barticle}
\bauthor{\bsnm{{Shen}}, \binits{Y.}},
\bauthor{\bsnm{{Ho}}, \binits{L.C.}}:
\bjtitle{\nat}
\bvolume{513},
\bfpage{210}
(\byear{2014}).
\arxivurl{1409.2887}.
doi:\doiurl{10.1038/nature13712}
\end{barticle}
\endbibitem

\bibitem[\protect\citeauthoryear{{Shen} and {Loeb}}{2010}]{shen10}
\begin{barticle}
\bauthor{\bsnm{{Shen}}, \binits{Y.}},
\bauthor{\bsnm{{Loeb}}, \binits{A.}}:
\bjtitle{\apj}
\bvolume{725},
\bfpage{249}
(\byear{2010}).
\arxivurl{0912.0541}.
doi:\doiurl{10.1088/0004-637X/725/1/249}
\end{barticle}
\endbibitem

\bibitem[\protect\citeauthoryear{{Shields} and {Bonning}}{2008}]{shields08}
\begin{barticle}
\bauthor{\bsnm{{Shields}}, \binits{G.A.}},
\bauthor{\bsnm{{Bonning}}, \binits{E.W.}}:
\bjtitle{\apj}
\bvolume{682},
\bfpage{758}
(\byear{2008}).
\arxivurl{0802.3873}.
doi:\doiurl{10.1086/589427}
\end{barticle}
\endbibitem

\bibitem[\protect\citeauthoryear{{Smith} et~al.}{2002}]{smith02}
\begin{barticle}
\bauthor{\bsnm{{Smith}}, \binits{J.E.}},
\bauthor{\bsnm{{Young}}, \binits{S.}},
\bauthor{\bsnm{{Robinson}}, \binits{A.}},
\bauthor{\bsnm{{Corbett}}, \binits{E.A.}},
\bauthor{\bsnm{{Giannuzzo}}, \binits{M.E.}},
\bauthor{\bsnm{{Axon}}, \binits{D.J.}},
\bauthor{\bsnm{{Hough}}, \binits{J.H.}}:
\bjtitle{\mnras}
\bvolume{335},
\bfpage{773}
(\byear{2002}).
\arxivurl{astro-ph/0205204}.
doi:\doiurl{10.1046/j.1365-8711.2002.05665.x}
\end{barticle}
\endbibitem

\bibitem[\protect\citeauthoryear{{Sobolev}}{1963}]{sobolev63}
\begin{bbook}
\bauthor{\bsnm{{Sobolev}}, \binits{V.V.}}:
\bbtitle{{A Treatise on Radiative Transfer.}},
(\byear{1963})
\end{bbook}
\endbibitem

\bibitem[\protect\citeauthoryear{{Tsalmantza} et~al.}{2011}]{tsalmantza11}
\begin{barticle}
\bauthor{\bsnm{{Tsalmantza}}, \binits{P.}},
\bauthor{\bsnm{{Decarli}}, \binits{R.}},
\bauthor{\bsnm{{Dotti}}, \binits{M.}},
\bauthor{\bsnm{{Hogg}}, \binits{D.W.}}:
\bjtitle{\apj}
\bvolume{738},
\bfpage{20}
(\byear{2011}).
\arxivurl{1106.1180}.
doi:\doiurl{10.1088/0004-637X/738/1/20}
\end{barticle}
\endbibitem

\bibitem[\protect\citeauthoryear{{Vestergaard} and
  {Peterson}}{2006}]{vestergaard06}
\begin{barticle}
\bauthor{\bsnm{{Vestergaard}}, \binits{M.}},
\bauthor{\bsnm{{Peterson}}, \binits{B.M.}}:
\bjtitle{\apj}
\bvolume{641},
\bfpage{689}
(\byear{2006}).
\arxivurl{astro-ph/0601303}.
doi:\doiurl{10.1086/500572}
\end{barticle}
\endbibitem

\bibitem[\protect\citeauthoryear{{Volonteri} et~al.}{2010}]{volonteri10}
\begin{barticle}
\bauthor{\bsnm{{Volonteri}}, \binits{M.}},
\bauthor{\bsnm{{G{\"u}ltekin}}, \binits{K.}},
\bauthor{\bsnm{{Dotti}}, \binits{M.}}:
\bjtitle{\mnras}
\bvolume{404},
\bfpage{2143}
(\byear{2010}).
\arxivurl{1001.1743}.
doi:\doiurl{10.1111/j.1365-2966.2010.16431.x}
\end{barticle}
\endbibitem

\bibitem[\protect\citeauthoryear{{Wang} et~al.}{2009}]{wang09}
\begin{barticle}
\bauthor{\bsnm{{Wang}}, \binits{J.-G.}},
\bauthor{\bsnm{{Dong}}, \binits{X.-B.}},
\bauthor{\bsnm{{Wang}}, \binits{T.-G.}},
\bauthor{\bsnm{{Ho}}, \binits{L.C.}},
\bauthor{\bsnm{{Yuan}}, \binits{W.}},
\bauthor{\bsnm{{Wang}}, \binits{H.}},
\bauthor{\bsnm{{Zhang}}, \binits{K.}},
\bauthor{\bsnm{{Zhang}}, \binits{S.}},
\bauthor{\bsnm{{Zhou}}, \binits{H.}}:
\bjtitle{\apj}
\bvolume{707},
\bfpage{1334}
(\byear{2009}).
\arxivurl{0910.2848}.
doi:\doiurl{10.1088/0004-637X/707/2/1334}
\end{barticle}
\endbibitem

\end{thebibliography}
\fussy

\end{document}